\documentclass[aps,pra,showpacs]{revtex4}

\newcommand{\nn}{\nonumber \\}
\newcommand{\tdb}{\tilde{\Delta}_b }
\newcommand{\tdf}{\tilde{\Delta}_f }

\newcommand{\rk}{\right)} 
\newcommand{\lk}{\left(} 
\newcommand{\rtk}{\right\} } 
\newcommand{\ltk}{\left\{ } 
\newcommand{\rdk}{\right] } 
\newcommand{\ldk}{\left[ } 

\usepackage{graphicx}

\begin{document}

%
\title{Density waves in quasi-one-dimensional 
atomic gas mixture of boson and two-component fermion}
%
%
\author{E. Nakano}
\affiliation{Yukawa Institute for Theoretical Physics, Kyoto University, 
             Sakyoku, Kyoto 606-8502, Japan} 
\author{H. Yabu}
\affiliation{Department of Physics, Tokyo Metropolitan University, 
            1-1 Minami-Ohsawa, Hachioji, Tokyo 192-0397, Japan}
\begin{abstract}
The density-wave structures are studied 
in quasi-one-dimensional atomic gas 
mixture of one-component bosons and two-component fermions 
using the mean-field approximation.
Owing to the Peierls instability in the quasi-one-dimensional fermion system, 
the ground state of the system shows 
the fermion density wave and the periodic Bose-Einstein condensation 
induced by the boson-fermion interatomic interaction. 
For the two-component fermions, 
two density waves appear in each component, 
and the phase difference between them distinguishes 
two types of ground states,  
the in-phase and the out-phase density-waves. 
In this paper, 
a self-consistent method in the mean-field approximation is presented 
to treat the density-wave states 
in boson-fermion mixture with two-component fermions.   
From the analysis of the effective potential and the interaction energies, 
the density-waves are shown to appear in the ground state, 
which are in-phase or out-phase 
depending on the strength of the inter-fermion interaction.  
It is also shown that the periodic Bose-Einstein condensate  
coexists with the in-phase density-wave of fermions, 
but, in the case of the out-phase one, only the uniform condensate appears. 
The phase diagram of the system is given for the effective coupling constants. 
\end{abstract}

\maketitle 

\section{Introduction} 

Since the experimental achievement 
of Bose-Einstein condensates (BEC) \cite{BECexp1, BECtext1} 
and Fermi-degenerate states \cite{FDexp1},  
the researches on the ultra-cold atomic gas have much stimulated 
the quantum many-body physics; 
it can create an artificial many-body system 
with the controllable the atom number, trapping potential geometry, 
the interatomic interactions et.al.   

     In recent years, 
the quantum-degenerate atomic gas of the boson-fermion mixture has been realized 
with the advanced cooling technique (the sympathetic cooling)  
\cite{BFMexp1, BFMexp2}, 
and a lot of experimental and theoretical studies 
have been done for it 
\cite{BFMtheory1,BFMtheory2,BFMtheory3,BFMtheory4,BFMtheory5,BFMtheory6,BFMtheory7,BFMtheory8,BFMtheory9}. 

The quasi-one-dimensional atomic (Q1D) gas of the boson-fermion mixture 
should open up many interesting possibilities in the quantum many-body system
\cite{MYS1}: 
the Tonk-Girardeau gas \cite{tonk1, tonk2}, 
and other regimes of trapped 1D bosons \cite{regime1} and  
of fermions \cite{regime2}. 
As discussed in \cite{MYS1}, 
the Q1D atomic gas can be realized, for example, 
in the axially-deformed harmonic oscillator (HO) trap 
with the axial and radial frequencies $\omega_{a,r}$, 
which satisfy $\omega_a \ll \omega_r$ and 
should be larger than the interatomic interaction energies and 
the healing length of the system \cite{MYS1, Das1, Pita1}. 
Experimentally, the quasi-one-dimensional systems has already been realized 
in the boson atomic gas \cite{BFMexp1, BEClow1, BEClow2, BEClow3}, 
where the BEC solitons propagating 
in the axial direction have been observed \cite{soliton1, soliton2, soliton3}. 

A particular property of the fermionic Q1D system is 
the occurrence of the density wave 
due to the Peierls instability
with the frequency of twice the Fermi wave number $2 k_F$.  
In \cite{MYS1}, 
we investigated the ground-state properties and 
the collective excitations 
in the mixture of one-component boson and one-component fermion, 
and showed that the fermion density wave appears in the ground state of the system, 
which is similar with the charge density wave (CDW) 
in the Q1D electron system \cite{peierl, Kittel, kagoshima}, 
and we also found that 
the BEC of the bosons becomes spatially periodic (periodic BEC).

In the present paper, 
we investigate the ground state properties 
of the Q1D atomic gas at zero-temperature ($T=0$)
of one-component boson and two-component fermion, 
which interact through the atomic collisions. 
In this system, 
two density waves can appear in each component of fermions 
in the ground state, 
and two types of the ground-state structure 
are distinguished with the relative phase between them,   
the in-phase and the out-phase density-waves (Fig.~1);  
\begin{figure}
\begin{center}
\includegraphics[height=6cm]{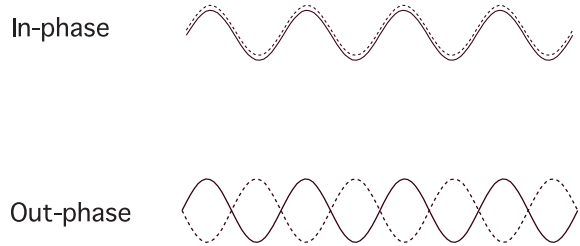}
\end{center}
\vspace{-1.5cm}
\caption{In-phase and out-phase density waves 
in fermion densities $\rho_\uparrow(z)$ (solid curve)
and $\rho_\downarrow(z)$ (dashed curve).}
\label{Fig1}
\end{figure}
the latter corresponds to the spin density wave (SDW) 
in the electron system \cite{kagoshima}. 
The density-wave structure in the ground state depends on 
the strength of the interatomic interactions;
especially important is the inter-component interaction of fermions, 
which is absent in the previous study \cite{MYS1}. 
Also, the structure of BEC, uniform or periodic,   
should be influenced by the fermion density-wave structure.

In Sec.~$\mbox{I\hspace{-.1em}I}$, 
we introduce a model Hamiltonian for the system, 
and propose a new method to treat the periodic order parameters 
for boson and fermion-density fields,  
which are proved to be self-consistent 
in the mean-field approximation, and 
construct the effective potential with this method. 
In Sec.~$\mbox{I\hspace{-.1em}I\hspace{-.1em}I}$,   
the density-wave structures in the ground state 
are classified by analytical evaluation of the effective potential 
for the cases of the repulsive/attractive fermion-fermion interactions;    
the results are summarized in the phase diagram 
for the effective coupling constants of the interatomic interactions.  
The final section is devoted to the summary of the paper. 

\section{Formulation} 

\subsection{Hamiltonian}

We consider the Q1D system of atomic gas mixture 
of one-component boson and two-component fermion at $T=0$. 
In this paper, 
we call two components of fermion ``up'' and ``down'', and  
denote them with the symbol $\uparrow$ and $\downarrow$. 
It should be noted that they should not only be the up and down components of spin 1/2 atom, 
but any two of hyperfine levels or different atomic species.

The masses of the bosons and the fermions are denoted 
by $m_b$ and $m_f$. 
In this paper, 
we consider the case of degenerate two-component fermion for simplicity, 
and assume that the up and down fermions have the same mass 
(also, we assume that the other properties, the interaction strengths et.al., 
are degenerate). 

The boson-fermion mixture is assumed to be trapped in the HO potentials 
with the same frequencies: 
\begin{equation}
     U_{b,f}=m_{b,f}\ltk \omega_r^2(x^2+y^2)+\omega_a^2z^2\rtk/2, 
\label{HOP}
\end{equation}
If the HO potentials are enough long in the z-direction, 
$\omega_r \gg \omega_a$, 
the system becomes Q1D. 
(For the other conditions in the realization of the Q1D system, 
see \cite{MYS1,Das1,Pita1}.)

In the Q1D system, 
the dynamical degrees of freedom of atoms are represented 
by the 1D field operators, 
$\phi(z)$ for boson and $\psi_{\uparrow, \downarrow}(z)$ for fermions. 
Using these field, 
the Hamiltonian of the system is given by
\begin{eqnarray}
     {\rm H} &=& \int dz \phi^\dagger(z) 
                 \left[ -\frac{\hbar^2}{2m_b}
                         \frac{{\rm d}^2}{{\rm d}z^2} 
                        +U_b'(z) -\mu_b  
                 \right] \phi(z)  \nn
              &&+ \int dz \sum_{s=\uparrow, \downarrow} 
                  \psi_s^\dagger(z) 
                  \left[ -\frac{\hbar^2}{2m_f}
                          \frac{{\rm d}^2}{{\rm d}z^2}
                         +U_f'(z) -\mu_f \right] \psi_s(z) \nn
              &&+ \int dz \phi^\dagger(z) 
                  \left[ \frac{g_{bb}}{2} \phi^\dagger(z) \phi(z)  
                        + g_{bf} \sum_{s=\uparrow, \downarrow} 
                                 \psi_s^\dagger(z) \psi_s(z) 
                  \right] \phi(z) \nn
              &&+ \int dz\; g_{ff} \psi_\uparrow^\dagger(z) 
                                 \psi_\uparrow(z) 
                                 \psi_\downarrow^\dagger(z)
                                 \psi_\downarrow(z),  
\label{Q1DH}
\end{eqnarray}
where $\mu_{b,f}$ are one-dimensional chemical potentials 
for the bosons and the fermions. 
The coupling constants $g_{bb,bf,ff}$ in eq.~(\ref{Q1DH}) 
represent the strengths of the boson-boson, boson-fermion 
and fermion-fermion interactions, respectively.
In the case of the large attractive boson-boson interaction ($g_{bb} \ll 0$), 
the BEC instability appears  
and competes with the Peierls instability. 
In the present paper, 
to concentrate on the density wave states, 
we consider the case of the repulsive boson-boson interaction ($g_{bb}>0$) only. 
For the Q1D atomic gas, 
the coupling constants are determined by the (3D) s-wave scattering lengths 
$a_{bb,bf,ff}$ of the atomic collisions \cite{pethick}:
$g_{bb}=2\hbar \omega_r a_{bb}$, 
$g_{bf}=2\hbar \omega_r a_{bf}$, and 
$g_{ff}=2\hbar \omega_r a_{ff}$, 
where $\omega_r$ is the radial frequency of the (3D) HO potential (\ref{HOP}). 
The $U_{b,f}'(z)$ in eq.~(\ref{Q1DH}) are the $z$-directional part 
in the original HO potential (\ref{HOP}), 
which are neglected in the present calculation. 
(About the effects of the HO potential, see \cite{MYS1}.)
 
\subsection{Density-wave ground state}

It is well known in the Q1D fermion system \cite{kagoshima,peierl,MYS1,Kittel} 
and also in the boson-fermion mixture \cite{MYS1} 
that the homogeneous state is unstable (Peierls instability)
for the formation of the density wave with the wave number $2k_F$.
To describe the density-wave states, 
we introduce the $2k_F$-periodic order parameters 
for the boson and fermion fields: 
\begin{eqnarray}
     \langle \phi(z) \rangle 
          &\equiv& \Phi(z) = b_0+2b_c \cos(2k_F z), 
\label{BD1} \\ 
     \rho_\downarrow(r) 
          &\equiv& \langle \psi_\downarrow^\dagger(z) 
                           \psi_\downarrow(z) \rangle
                =f_0 +2f_c \cos(2k_F z+\varphi)
                     +2f_b \cos(2k_F z), 
\label{FD1} \\ 
     \rho_\uparrow(r) 
          &\equiv& \langle \psi_\uparrow^\dagger(z) 
                           \psi_\uparrow(z) \rangle 
                =f_0 +2f_c \cos(2k_F z +\varphi+\theta)
                +2f_b \cos(2k_F z), 
\label{FD2}
\end{eqnarray}
where $\varphi$ is the phase-difference angles 
between boson- and fermion-density waves, 
and $\theta$ is that between the up- and down-fermion density waves;  
$\theta=0 ~(\pi)$ corresponds to the in-phase (out-phase) density wave of fermions (Fig.~1). 
The $b_c$ and $f_c$ in eq.~(\ref{FD2}) are the amplitudes of the boson
and fermion density waves, 
and the third components with the amplitude $f_b$ in $\rho_{\downarrow,\uparrow}$ 
have to be added to reflect the fermion-density waves 
induced by the periodic BEC 
(the term with the amplitude $b_c$ in the boson order parameter).  
Thus, the amplitude $f_b$ will be shown 
to be proportional to the $b_c$ at the end of this section.   
The constants $b_0$ and $f_0$ represent the homogeneous terms 
in the boson and fermion densities, 
which are determined by the boson and fermion average densities, 
$n_b$ and $n_f$ after the spatial integration of the local densities: 
\begin{equation}
     n_b=b_0^2+2b_c^2, \quad 
     n_f=n_{f, \uparrow}=n_{f, \downarrow}=f_0. 
\label{density}
\end{equation}
In the present study, 
we do not consider the case 
where the spin-polarized states 
with $n_{f, \uparrow}-n_{f, \downarrow} \neq 0$ 
become stable; 
in the cases of the very large fermion density 
(or the very strong fermion-fermion interaction), 
such a polarized state might be stable 
(the collective ferromagnetic states) \cite{SpinPol}. 

Using eqs.~(\ref{BD1}-\ref{FD2}) for the Hamiltonian (\ref{Q1DH}), 
we obtain the mean-field Hamiltonian density:  
\begin{eqnarray}
     {\mathcal H}_{MF}
          ={\mathcal H}_F 
          +{\mathcal H}_B, 
\label{MFH0}
\end{eqnarray}
where the fermion part ${\mathcal H}_F$ 
includes the contributions from the fermion kinetic term 
and the fermion-fermion and boson-fermion interaction terms, 
and the boson part ${\mathcal H}_B$ 
comes from the boson kinetic and boson-boson interaction terms.
Using the periodic regularization with the spatial length $L$ 
and taking the limit $L\rightarrow \infty$,  
the fermion part for the momentum $-2k_F < k < 0$, 
denoted by ${\mathcal H}_F^-$, becomes
\begin{eqnarray}
     {\mathcal H}_F^-&=& \int \frac{dk}{2\pi}
                             \left( \begin{array}{c}
                                    \psi_\uparrow(k) \\ 
                                    \psi_\uparrow(k+2k_F)
                                    \end{array}
                             \right)^\dagger
   \left[ \begin{array}{c c}
     \epsilon(k) &   \Delta_b +\Delta_b' +e^{i\varphi} \Delta_f   \\
     \Delta_b^* +{\Delta_b'}^* +e^{-i\varphi} \Delta_f^*    &  \epsilon(k+2k_F)
   \end{array} \right]
                             \left( \begin{array}{c}
                                    \psi_\uparrow(k) \\
                                    \psi_\uparrow(k+2k_F)
                                    \end{array}\right) \nn
                         &+& \int \frac{dk}{2\pi}
                             \left( \begin{array}{c}
                                    \psi_\downarrow \\ 
                                    \psi_\downarrow(k+2k_F)
                                    \end{array} \right)^\dagger
  \left[\begin{array}{c c}
    \epsilon(k) &   \Delta_b +\Delta_b' +e^{i(\theta+\varphi)} \Delta_f  \\
    \Delta_b^* +{\Delta_b'}^* +e^{-i(\theta+\varphi)} \Delta_f^* & \epsilon(k+2k_F)
  \end{array} \right]
                             \left( \begin{array}{c}
                                    \psi_\downarrow(k) \\
                                    \psi_\downarrow(k+2k_F)
                                    \end{array} \right) \nn
                         &-& \frac{g_{ff}}{2} f_0^2 
                           -\frac{1}{g_{ff}} 
                            \left[ \Delta_f^2\cos(\theta) 
                                  +2\Delta_f \Delta_b' 
                                    \cos\left(\frac{\theta}{2} \right) 
                                    \cos\left(\varphi+\frac{\theta}{2} \right) 
                                  +\Delta_b'^2 \right].
\label{MFH1}
\end{eqnarray}
The fermion part, ${\mathcal H}_F^+$, 
corresponding to the positive momentum $0<k<2k_F$,
is obtained
by the replacement $k_F \rightarrow -k_F$ in eq.~(\ref{MFH1}), 
and the total fermion part becomes   
${\mathcal H}_F={\mathcal H}_F^-+ {\mathcal H}_F^+$. 
The $\epsilon(k)$ is the single-particle energy 
$\epsilon(k) = \hbar^2 k^2/2 m_f-\mu_f +g_{ff} f_0 +g_{bf}n_b$, 
and the gap functions $\Delta$ are defined by 
\begin{equation}
     \Delta_b \equiv 2b_0 b_c g_{bf},  \quad 
     \Delta_f \equiv f_c g_{ff},       \quad
     \Delta_b' \equiv f_b g_{ff}.
\label{gapf}
\end{equation}   

Using the similar technique with the Bogoliubov transformation, 
the ${\mathcal H}_F$ is diagonalized 
(for the detailed calculation, see Appendix A):
\begin{eqnarray}
     {\mathcal H}_F^- &=&  \int \frac{dk}{2\pi} 
                            \sum_{s={\uparrow, \downarrow}} 
                                 \left[ E^-_s(k) {\alpha^-_s}^\dagger(k) 
                                                 \alpha^-_s(k)
                                       +E^+_s(k) {\alpha^+_s}^\dagger(k) 
                                                 \alpha^+_s(k) \right]      \nn
                    &-& \frac{g_{ff}}{2} f_0^2 
                          -\frac{1}{g_{ff}} 
                           \left[ \Delta_f^2\cos(\theta) 
                                 +2\Delta_f \Delta_b' 
                                   \cos\left(\frac{\theta}{2} \right) 
                                   \cos\left(\varphi+\frac{\theta}{2} \right) 
                                 +\Delta_b'^2 \right], 
\label{fermi1}
\end{eqnarray}
where $\alpha^\pm_{\uparrow, \downarrow}(k)$ are the annihilation operators 
for the quasi-particle with the momentum $-2k_F < k < 0$
in the energies: 
\begin{eqnarray}
     E^\pm_\uparrow(k) &=& \left[ \epsilon(k)+v_F(k+k_F) \right]
                        \pm \sqrt{ v_F^2(k+k_F)^2 
                        +|\Delta_{\uparrow}|^2 },  
\label{qpe1} \\
     E^\pm_\downarrow(k) &=& \left[ \epsilon(k)+v_F(k+k_F) \right]
                          \pm \sqrt{ v_F^2(k+k_F)^2
                          +|\Delta_{\downarrow}|^2},  
\label{qpe2} 
\end{eqnarray}
where $v_F$ is the Fermi velocity. 
The quasi-particle energies for the momentum $0<k<2k_F$ are obtained by replacing 
$k_F, v_F \rightarrow -k_F, -v_F$ in eqs.~(\ref{qpe1},\ref{qpe2}).   
The effective gap functions $\Delta_{\uparrow, \downarrow}$ in eqs.~(\ref{qpe1},\ref{qpe2})
are defined by
\begin{eqnarray}
     |\Delta_{\uparrow}|^2 
          &\equiv& \left| \Delta_b +\Delta_b' 
                         + e^{i\varphi}\Delta_f \right|^2 
                 =(\Delta_b +\Delta_b')^2 
                 +2 (\Delta_b+\Delta_b') \Delta_f \cos(\varphi) 
                 +\Delta_f^2, 
\label{egf1}\\
     |\Delta_{\downarrow}|^2 
          &\equiv& \left| \Delta_b +\Delta_b' 
                         + e^{i(\varphi+\theta)}\Delta_f \right|^2 
                =(\Delta_b+\Delta_b')^2 
                +2 (\Delta_b +\Delta_b') \Delta_f \cos(\theta+\varphi) 
                +\Delta_f^2. 
\label{egf2}
\end{eqnarray}

\begin{figure}
\begin{center}
\includegraphics[height=5cm]{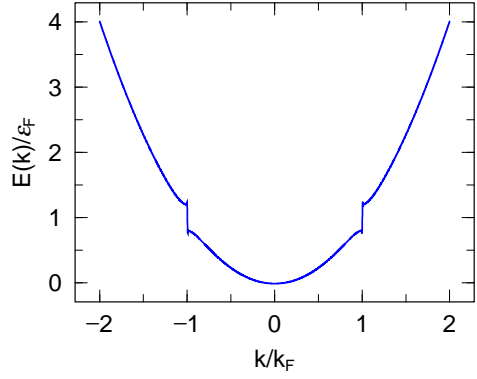}
\end{center}
\caption{Illustration of the quasi-particle spectrum, 
$E_{\uparrow, \downarrow}^{(\pm)}(k)$. 
We scaled quantities by the fermi energy and fermi momentum, 
and set $\Delta_{\uparrow, \downarrow}/2\epsilon_F=0.1$. }
\label{level1}
\end{figure}

In Fig.~\ref{level1}, we show the quasi-particle energy in eqs.~(\ref{qpe1},\ref{qpe2}). 
It should be noted that the energy gaps appear at $k =\pm k_F$ 
in the quasi-particle spectrum. 

In the density-wave ground state, 
the up- or down-fermions occupy 
the lower branch of the quasi-particle spectrum 
so that the energy density of fermions in this state 
is given by the summation of $E^-_{\uparrow (\downarrow)}(k)$ 
in eqs.~(\ref{qpe1},\ref{qpe2})
up to the Fermi momentum $k_F$:
\begin{eqnarray}
     \langle {\mathcal H}_F\rangle   
          &&= \sum_{s=\uparrow, \downarrow} 
              \int \frac{dk}{2\pi} 
              \left[ E^-_s(k,-k_F) \theta(k_F-k)\theta(k) 
                    +E^-_s(k,k_F) \theta(k_F+k)\theta(-k) \right]  \nn 
          &&\quad 
             - g_{ff} f_0^2 
             -\frac{2}{g_{ff}} \left[ \Delta_f^2\cos(\theta) 
                                     +2\Delta_f \Delta_b' 
                                       \cos\left(\frac{\theta}{2} \right) 
                                       \cos\left(\varphi+\frac{\theta}{2} \right) 
                                     +\Delta_b'^2 \right] \nn 
          &&= n_f \left[ \frac{4\epsilon_F}{3} 
                         +g_{ff} f_0 
                         -\mu_f 
                         + g_{bf} n_b \right] \nn
          &&\quad
             -\frac{n_f}{2} \sum_{s=\uparrow, \downarrow} 
              \left[ \sqrt{4\epsilon_F^2 +\Delta_s^2}
                                   +\frac{\Delta_s^2}{2 \epsilon_F} 
                     \log{\frac{2 \epsilon_F
                                +\sqrt{4 \epsilon_F^2 
                                       +\Delta_s^2}}{\Delta_s}} \right] \nn 
          &&\quad
            - g_{ff}f_0^2 
             -\frac{2}{g_{ff}} 
              \left[ \Delta_f^2\cos(\theta) 
                   +2\Delta_f \Delta_b' 
                     \cos\left(\frac{\theta}{2} \right) 
                     \cos\left(\varphi+\frac{\theta}{2} \right) 
                   +\Delta_b'^2 \right], 
\label{fmfe}
\end{eqnarray}
where we have used 
$\langle {\mathcal H}_F\rangle= 
\langle {\mathcal H}_F^-\rangle+ \langle{\mathcal H}_F^+\rangle
=2 \langle{\mathcal H}_F^-\rangle$.

Now we turn to the evaluation of the mean-field energy 
for the boson part ${\mathcal H}_B$ in eq.~(\ref{MFH1}). 
Substituting eq.~(\ref{BD1}) into the boson part,  
we obtain the boson energy density up to $O(b_c^2)$:  
\begin{eqnarray}
     \langle {\mathcal H}_B \rangle  
           &=& 2 \left[ \epsilon_b(2k_F) 
                       +6 \frac{g_{bb}}{2}b_0^2 \right] b_c^2 
              +\frac{g_{bb}}{2}b_0^4 
              +6 \frac{g_{bb}}{2}  b_c^4 \nn 
      &\simeq& 2 \left[ \epsilon_b(2k_F)
                       +5 \frac{g_{bb}}{2}b_0^2 \right]b_c^2 
             +\frac{g_{bb}}{2} b_0^2 n_b \nn  
            &=& 2 \epsilon_b(2k_F) b_c^2 
               +10 \frac{g_{bb}}{2} \frac{\Delta_b^2}{4g_{bf}^2} 
               +\frac{g_{bb}}{2} b_0^2 n_b \nn
      &\simeq& \frac{g_{bb}}{2} n_b^2 
              +\left( \frac{k_F^2}{m_b n_b} +g_{bb} \right) 
               \left( \frac{\Delta_b}{g_{bf}} \right)^2. 
\label{bmfe}
\end{eqnarray}
In the derivation of the last line, 
we have used the solution of 
$n_b=b_0^2+2b_c^2$ (the average boson density) 
and $\Delta_b \equiv 2b_0 b_c g_{bf}$: 
\begin{eqnarray} 
     b_0^2 &=& \frac{1}{2}
               \left[ n_b 
                     +\sqrt{n_b^2 -2 \left(\frac{\Delta_b}{g_{bf}} \right)^2} 
               \right], \\
     2b_c^2 &=& \frac{1}{2}
                \left[ n_b
                      -\sqrt{n_b^2 -2 \left(\frac{\Delta_b}{g_{bf}} \right)^2} 
                \right], \\
     n_b^2 &\ge& 2 \left(\frac{\Delta_b}{g_{bf}} \right)^2,  
\end{eqnarray}
where we have assumed the small modulation 
in the boson density $b_0^2 \gg 2 b_c^2$, 
and taken the solution satisfying $b_0^2=n_b$ and $2b_c^2=0$ at $\Delta_b=0$.   

Thus the total energy density is given by the sum of eqs.~(\ref{fmfe},\ref{bmfe}) :
\begin{eqnarray}
     \mathcal{E}_{tot}(\Delta_b, \Delta_f, \Delta_b'; \varphi, \theta) 
          &\equiv& \langle {\mathcal H}_{MF}\rangle 
                  =\langle {\mathcal H}_F\rangle
                  +\langle {\mathcal H}_B\rangle. 
\label{EVH} 
\end{eqnarray} 
Explicit representation of the energy difference 
$\tilde{\mathcal{E}}_{tot}(\Delta_b, \Delta_f, \Delta_b')$
between the density-wave state and the normal state 
($\Delta_b =\Delta_f =\Delta_b =0$) 
is given by
\begin{eqnarray}
     && \tilde{\mathcal{E}}_{tot}(\Delta_b, \Delta_f, \Delta_b')
          \equiv \mathcal{E}_{tot}(\Delta_b, \Delta_f, \Delta_b')
                -\mathcal{E}_{tot}(0, 0, 0)   \nn
     &&\qquad 
          = -\sum_{s=\uparrow, \downarrow} 
             \frac{n_f}{2} 
             \left[ \sqrt{4\epsilon_F^2 +\Delta_s^2}
                   +\frac{\Delta_s^2}{2 \epsilon_F} 
                    \log \frac{2 \epsilon_F 
                               +\sqrt{4 \epsilon_F^2 
                                      +\Delta_s^2} 
                             }{\Delta_s}
             \right] 
           +\left( \frac{k_F^2}{m_b n_b} +g_{bb} \right)  
            \left(\frac{\Delta_b}{g_{bf}} \right)^2      \nn
     &&\qquad\quad
          +2 n_f \epsilon_F 
          -\frac{2}{g_{ff}} 
           \left[ \Delta_f^2\cos\theta 
                 +2 \Delta_f \Delta_b' 
                  \cos\left(\frac{\theta}{2} \right) 
                  \cos\left(\varphi+\frac{\theta}{2} \right) 
                 +\Delta_b'^2 \right] 
\label{ED1} \\
     &&\qquad 
          \simeq 2 n_f \epsilon_F 
                 \Bigg\{ -\sum_{s=\uparrow, \downarrow} 
                          \frac{1}{4} 
                 \left[ 1 
                       -2\log \left( \frac{\Delta_s 
                                          }{4\epsilon_F} 
                             \right) 
                 \right] 
                 \left( \frac{\Delta_s}{2\epsilon_F} \right)^2  
                +\frac{1}{\zeta} 
                 \left( \frac{\Delta_b}{2\epsilon_F} \right)^2 
                     \nn
     &&\qquad\qquad\qquad
          -\frac{2}{\xi} 
           \left[ \left( \frac{\Delta_f}{2\epsilon_F} \right)^2 
                  \cos\theta 
                 +2 \frac{\Delta_f \Delta_b'}{(2\epsilon_F)^2}
                    \cos\left(\frac{\theta}{2} \right) 
                    \cos\left(\varphi+\frac{\theta}{2} \right) 
                              +\left(\frac{\Delta_b'}{2\epsilon_F}\right)^2 
           \right] \Bigg\},  
\label{ED3}
\end{eqnarray}
where dimensionless parameters $\zeta$ and $\xi$ 
(effective coupling constants) have been introduced:
\begin{eqnarray} 
     \zeta &\equiv& \frac{g_{bf}^2}{\pi \hbar g_{bb} v_F}
                  \left[ 1 +\left( \frac{v_F m_f}{v_B m_b} \right)^2  \right]^{-1},  
\label{vzeta}\\
     \xi &\equiv& \frac{g_{ff}}{\pi \hbar v_F}, 
\label{vxi}
\end{eqnarray}
and $v_B=\sqrt{g_{bb} n_b/m_b}$ is the bosonic velocity.  

The stable ground state of the system is determined 
from the minimum point of eq.~(\ref{ED3}). 

\subsection{Consistency of the order parameters in mean-field approximation}

Here we check the consistency 
of the Fermion density ansatz~(\ref{FD1}, \ref{FD2}) 
in the mean-field approximation,  
where the fermion densities $\rho_{\uparrow, \downarrow}(z)$ 
are represented by
\begin{eqnarray} 
     \rho_s(z) &\equiv& \langle \psi^\dagger_s(z)\psi_s(z) \rangle  \nn
                    &=& \int \frac{dp}{2\pi} \int \frac{dk}{2\pi} 
                        \langle \psi^\dagger_s(p)\psi_s(k) \rangle 
                        e^{i(p-k)z} \nn 
                    &=& \int \frac{dk}{2\pi} 
                             \langle \psi^\dagger_s(k) \psi_s(k) \rangle \nn
                    && +\int \frac{dk}{2\pi} 
                             \langle \psi^\dagger_s(k+2k_F) \psi_s(k) \rangle 
                             e^{i2k_F z} 
                       +{\rm C.C.} ,  \qquad 
                    (s =\uparrow, \downarrow). 
\label{dex}
\end{eqnarray}
In the quasi-particle representation (shown in Appendix A), 
the expectation values appeared in eq.~(\ref{dex}) become
\begin{eqnarray} 
     \int \frac{dk}{2\pi} \langle \psi^\dagger_s(k)\psi_s(k) \rangle 
          &=& n_f, 
\label{dexqp1}\\
     \int \frac{dk}{2\pi} \langle \psi^\dagger_s(k+2k_F) \psi_s(k) \rangle 
          &=& 2 \int_{-k_F}^0 \frac{dk}{2\pi} 
                \frac{-\Delta_s}{2 \sqrt{ \ltk \epsilon(k) 
                                         -\epsilon(k+2k_F) \rtk^2 
                                         +|\Delta_s|^2 }}. 
\label{dexqp2}
\end{eqnarray}
Using the above equations in eq.~(\ref{dex}), 
we obtain the quasi-particle representation of the fermion densities:
\begin{eqnarray} 
     \rho_\uparrow(z) = n_f 
                       &+& 2 \int_{-k_F}^0 \frac{dk}{2\pi} 
         \frac{-(\Delta_b+\Delta_b')}{\sqrt{ \ltk\epsilon(k)
                                            -\epsilon(k+2k_F)\rtk^2
                                            +|\Delta_\uparrow|^2 }} 
                           \cos(2k_F z) \nn 
                      &+& 2 \int_{-k_F}^0 \frac{dk}{2\pi} 
        \frac{-\Delta_f}{\sqrt{ \ltk \epsilon(k) 
                               -\epsilon(k+2k_F) \rtk^2 
                               +|\Delta_\uparrow|^2 }} 
                           \cos(2k_F z+\varphi),   
\label{FD1re} \\ 
{} \nn
      \rho_\downarrow(z)  = n_f 
                        &+& 2 \int_{-k_F}^0 \frac{dk}{2\pi} 
         \frac{-(\Delta_b +\Delta_b')}{\sqrt{ \ltk\epsilon(k)
                                             -\epsilon(k+2k_F)\rtk^2
                                             +|\Delta_\downarrow|^2 }} 
         \cos(2k_F z) \nn 
                         &+& 2 \int_{-k_F}^0 \frac{dk}{2\pi} 
         \frac{-\Delta_f}{\sqrt{ \ltk\epsilon(k) 
                                -\epsilon(k+2k_F)\rtk^2
                                +|\Delta_\downarrow|^2 }} 
                               \cos(2k_F z+\theta+\varphi).    
\label{FD2re}
\end{eqnarray}
Here, the densities $\rho_{\uparrow, \downarrow}(z)$ above show  
the same $z$-dependence with those in eqs.~(\ref{FD1}, \ref{FD2});
it proved the consistency of the present method. 

In the present paper, 
we consider the case of $n_{f,\uparrow}=n_{f,\downarrow}$, i.e., no polarization, 
and have assumed the same amplitudes in ansatz (\ref{FD1},\ref{FD2}),  
so that we can put $|\Delta_{\downarrow}|=|\Delta_{\uparrow}|$. 
From eqs.~(\ref{egf1}, \ref{egf2}) 
for the gap functions $\Delta_{\uparrow, \downarrow}$, 
we obtain the relation between the phase-difference angles: 
$\cos(\varphi)=\cos(\theta+\varphi)$. 


Furthermore, 
we can derive the relation among the gap functions  
at the stationary points of the mean-field energy (\ref{EVH}). 
From the gap equations derived in Appendix~B, 
we obtain    
\begin{equation}
     \Delta_b'=-\frac{\xi}{2 \zeta} \Delta_b
                 -\Delta_f \cos\left(\frac{\theta}{2} \right) 
                           \cos\left(\varphi+\frac{\theta}{2} \right).
\label{delbp0}
\end{equation}
Using the definition of the gap functions in (\ref{gapf}), 
it can be written as the relation of the density-wave amplitudes:
\begin{equation}
     f_b = -\frac{\Delta_b}{2 G_{bf}} 
             -f_c \cos\left(\frac{\theta}{2} \right) 
                  \cos\left(\varphi+\frac{\theta}{2} \right), 
\label{delbp}
\end{equation}
where the effective boson-fermion interaction parameter $G_{bf}$ is defined by 
\begin{equation}
     G_{bf} \equiv g_{bf}^{2} \left( \frac{k_F^2}{m_b n_b} +g_{bb} \right)^{-1}.
\label{effbf}
\end{equation}
In the present paper, we consider the case of $g_{bb}>0$, 
so that the parameter $G_{bf}$ is positive ($G_{bf}>0$).  
The relation (\ref{delbp0}) shows 
that $\Delta_b'$ is related to
a linear combination of $\Delta_b$ and $\Delta_f$ in the ground state.  
In the next section, 
we will show that $\Delta_b'$ is proportional only 
to the boson amplitude $\Delta_b$ 
after fixing the phase-difference angles in the ground state.  

\subsection{Interaction energy and phase differences} 

To determine the ground state, 
we should fix the phase-difference angles $\varphi$ and $\theta$ 
so as to gain the largest interaction energy for the fixed gap functions. 
For this purpose, 
we reevaluate the $\varphi$- and $\theta$-dependent terms in (\ref{ED3}), 
included in the boson-fermion and fermion-fermion interaction terms in eq.~(\ref{Q1DH}),
in the mean-field approximation. 

Using the parametrization for order parameters (\ref{BD1},\ref{FD1},\ref{FD2}), 
the boson-fermion interaction term becomes  
\begin{eqnarray}  
     && g_{bf} \int dz  \sum_{s=\uparrow, \downarrow} 
     \phi^\dagger(z) \phi(z) \psi_s^\dagger(z) \psi_s(z) \nn
     &&\Rightarrow g_{bf} \int_0^L \frac{dz}{L} 
                     \sum_s \Phi^\dagger(z) \Phi(z) 
     \langle \psi_s^\dagger(z) \psi_s(z) \rangle |_{L\rightarrow \infty} \nn
                 &&\qquad= 2 g_{bf} 
                    \left[ n_b f_0 
                          +4 b_0 b_c f_c 
                           \cos\left(\frac{\theta}{2} \right)
                           \cos\left(\varphi+\frac{\theta}{2} \right) 
                          +4b_0 b_c f_b \right] \\
                 &&\qquad= 2 g_{bf} n_b n_f -2\frac{\Delta_b^2}{G_{bf}},  
\label{bfie}
\end{eqnarray}
where, in the derivation of the last line, 
we have used the relation (\ref{delbp}), 
which is valid for the ground state. 
Eq.~(\ref{bfie}) shows that the boson-fermion interaction term 
is independent of the phase-difference angles in the ground state, 
and plays no role in the phase difference determination.

In the same way, 
we reevaluate  
the fermion-fermion interaction term: 
\begin{eqnarray} 
     && g_{ff}\int dz\; \psi_\uparrow^\dagger(z) 
                                 \psi_\uparrow(z) 
                                 \psi_\downarrow^\dagger(z)
                                 \psi_\downarrow(z) \nn
    &&\Rightarrow g_{ff} \int_0^L \frac{dz}{L}  
             \langle \psi_\uparrow^\dagger(z) \psi_\uparrow(z) \rangle 
     \langle \psi_\downarrow^\dagger(z) \psi_\downarrow(z) \rangle |_{L\rightarrow \infty} \nn 
            &&\qquad= 2 g_{ff} \ldk \frac{f_0^2}{2}
                             +\ltk f_c \cos\left(\frac{\theta}{2} \right)
                                       \cos\left(\varphi+\frac{\theta}{2} \right) 
                                  +f_b \rtk^2 \right. \nn
            &&\qquad\qquad\qquad
                        + f_c^2 \left. \ltk -\cos\left(\frac{\theta}{2} \right)^2 
                         \cos\left(\varphi+\frac{\theta}{2} \right)^2 
                        +\cos\theta \rtk \rdk \nn
            &&\qquad= 2 g_{ff}\ldk \frac{n_f^2}{2} 
                            +\frac{\Delta_b^2}{4G_{bf}^2} \right. \nn
            &&\qquad\qquad\qquad
                    +f_c^2 \left. \ltk \cos\left(\frac{\theta}{2} \right)^2 
                                        \sin\left(\varphi+\frac{\theta}{2} \right)^2
                            -\sin\left(\frac{\theta}{2} \right)^2 \rtk \rdk. 
\label{ffie}
\end{eqnarray}
The phase-difference angles $\varphi$ and $\theta$ should be determined 
to make the energy gain largest in eq.~(\ref{ffie}) 
for fixed values of the coupling constant $g_{ff}$. 
Therein,  
the $(\theta,\phi)$-dependent part appears only in the second term, 
so that the phase-difference angles are determined 
as the minimum point of eq.~(\ref{ffie}); 
$(\theta,\varphi)=(0,\pi/2)$ for $g_{ff} <0$,
and $(\theta,\varphi)=(\pi,\pi/2)$ for $g_{ff} >0$. 
It should be noted that the results are consistent with the condition 
$\cos(\varphi)=\cos(\theta+\varphi)$ 
obtained in previous subsection.

\section{Classification of the density-wave structures in the ground state}

Now, we analyze the minimum of the mean-field energy (\ref{ED3}) 
with the phase difference conditions obtained in the last part of sec.~II, 
to determine the stable state of the system, 
and show the complete classification of the ground state
for the dimensionless coupling constants $\zeta$ and $\xi$. 

\subsection{Repulsive fermion-fermion interaction $g_{ff}>0$} 

In the case of $g_{ff} >0$, 
using $(\theta, \varphi)=(\pi, \pi/2)$ 
for the phase difference angles
in the ground state in eq.~(\ref{delbp0}), 
we obtain 
$\Delta_b'=-\frac{g_{ff}}{2 G_{bf}}\Delta_b \lk=-\frac{\xi}{2\zeta}\Delta_b\rk$.
Substituting them into eq.~(\ref{ED3}), 
we obtain the total energy density up to the order of $O(\Delta^2)$:
\begin{eqnarray}
     \tilde{\mathcal{E}}_{tot} = 2 n_f \epsilon_F\ltk -\frac{1}{2} 
          \left[ 1 -\log{\left(\frac{\tilde{\Delta}_B^2+\tdf^2}{4}\right)} \right] 
         \lk \tilde{\Delta}_B^2+\tdf^2 \rk  
        +\frac{2}{2\zeta-\xi}\tilde{\Delta}_B^2
        +\frac{2}{\xi} \tdf^2 \rtk, 
\label{eED1}
\end{eqnarray}
where $\tilde{\Delta}_B \equiv \lk 1-\frac{\xi}{2\zeta} \rk \Delta_b/(2\epsilon_F)$, 
and $\tdf \equiv \Delta_f/(2\epsilon_F)$. 

\begin{figure}
\begin{center}
\includegraphics[height=13.5cm, angle=90]{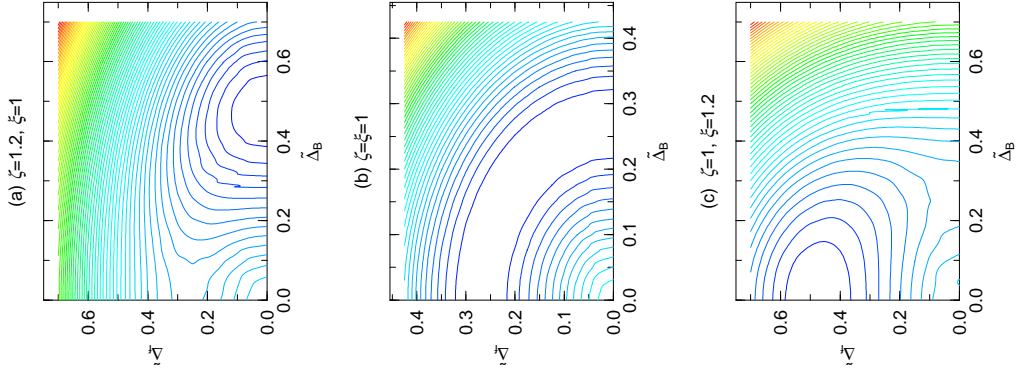}
\end{center}
\caption{Contour plots of $\tilde{\mathcal{E}}_{tot}$ 
in $\tilde{\Delta}_B$-$\tilde{\Delta}_f$ plane
for the repulsive Fermion-Fermion interaction $g_{ff}>0$. 
Three plots are for $\zeta>\xi$ (a), 
$\zeta=\xi$ (b) 
and $\zeta<\xi$ (c). }
\label{Figa}
\end{figure}

In Fig.~\ref{Figa}, 
the contour plots of the energy density (\ref{eED1}) 
are shown in $\tilde{\Delta}_B$-$\tilde{\Delta}_f$ plane
for typical values of the coupling constants $\zeta$ and $\xi$. 

Since the first term in eq.~(\ref{eED1}) is symmetric 
with respect to the order parameters $\tilde{\Delta}_{B,f}$,
the position of the minimum point of eq.~(\ref{eED1}) 
is determined by the last two terms; 
the results are summarized as 

\begin{itemize}
\item{$0<\xi<\zeta$ \qquad (Fig.~\ref{Figa}a)}, \qquad
$\lk \tdb,\tdf \rk 
= \lk 2\zeta \frac{2}{2\zeta-\xi}e^{\frac{-2}{2\zeta-\xi}}, 0 \rk$.   

In this case,  
the up- and down-fermions form in-phase density wave:  
\begin{equation}
     \rho_\downarrow(z)  = \rho_\uparrow(z)
                         =n_f \ldk 1- \frac{\Delta_b}{2 \epsilon_F \zeta} \cos(2k_F z) \rdk,
\label{fden1}
\end{equation}
and the bosons make a periodic BEC: 
\begin{equation}
     n_b(z) \simeq n_b \ldk 1 +\frac{\Delta_b}{ g_{bf} n_b} \cos(2k_F z) \rdk, 
\label{bden1}
\end{equation}
which is analogous to charge-density waves (CDW) of electrons 
in low dimensional conductors \cite{kagoshima, CDW2} 
due to the Peierls instability. 
We here mention the roles which the sign of $g_{bf}$ plays in the density waves. 
Since the amplitude of the periodic BEC in (\ref{bden1}) is always positive 
by definition,   
the phase difference between boson and fermion density waves 
is determined by the sign of the amplitude of the fermion density wave 
in (\ref{fden1}), 
which is proportional to the sign of $g_{bf}$:   
attractive (repulsive) boson-fermion interaction 
for the in-phase (out of phase).     
\item{$\zeta=\xi$ \qquad (Fig.~\ref{Figa}b)}, \qquad
$\tilde{\Delta}_b^2/4+\tilde{\Delta}_f^2=4e^{\frac{-4}{\xi}}$. 

The above equation shows the existence of the rotation-like ``hidden'' symmetry 
in the ground state,
which mixes between the bosonic and fermionic particle-hole pair order parameters. 
It also suggests the occurrence of the spontaneous symmetry breaking 
in the ground state 
and the existence of the Nambu-Goldstone mode
as a low-lying excitation mode.  
\item{$\zeta<\xi<2\zeta$ \qquad (Fig.~\ref{Figa}c) \ and\  $\xi \leq 2\zeta$}, \qquad
$\lk \tdb,\tdf \rk = \lk 0, 2e^{\frac{-2}{\xi}} \rk$. 

In this case,   
the fermion-fermion interaction induces the out-phase fermion-density waves:
\begin{equation}
     \rho_{\uparrow, \downarrow}(z) 
          =n_f\ldk 1 \pm  \frac{\Delta_f}{\epsilon_F \xi} \sin(2k_F z) \rdk,
\end{equation}
which forms the out-phase density wave 
in the two-component fermion system, analogous to the SDW \cite{ove, sdw1}. 

When $2\zeta\leq\xi$, 
we need a special care to treat the boson density wave, 
because the coefficient of $\tdb^2$ term in (\ref{eED1}) becomes negative, 
which can be traced back to the potential energy of the gap functions.  
Assembling the boson-fermion and fermion-fermion interaction terms (\ref{bfie}, \ref{ffie})
and the boson energy (\ref{bmfe}),  
the potential energy up to the constant, $H_{pot}$, becomes 
\begin{eqnarray}
         H_{pot}=  -\frac{1}{G_{bf}} \lk 1
          -\frac{g_{ff}}{2G_{bf}} \rk \Delta_b^2
          -\frac{2}{g_{ff}}\Delta_f^2, 
\end{eqnarray}
which should be negative definite for any value of the gap functions 
to gain the condensation energy.  
Thus, it gives $\Delta_b \equiv 0$ for $2G_{bf}\le g_{ff}$ (or equivalently for $2\zeta\le \xi$).  


\end{itemize}

\subsection{Attractive fermion-fermion interaction $g_{ff}<0$} 

\begin{figure}
\begin{center}
\includegraphics[height=4.5cm]{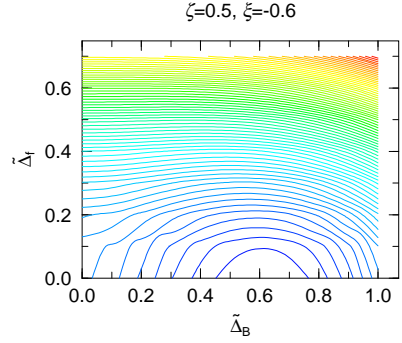}
\end{center}
\caption{Contour plots of $\tilde{\mathcal{E}}_{tot}$ 
in $\tilde{\Delta}_B$-$\tilde{\Delta}_f$ plane
for the attractive fermion-fermion interaction $g_{ff}<0$.}
\label{Figc}
\end{figure}

In the case of $g_{ff}<0$, 
using the phase difference angles $(\theta, \varphi)=(0, \pi/2)$ 
the relation $\tdb'=-\frac{\xi}{2\zeta}\tdb$ is obtained
from eq.~(\ref{delbp0}) in the ground state.
The total energy density (\ref{ED3}) becomes 
\begin{eqnarray}
     \tilde{\mathcal{E}}_{tot} =2 n_f \epsilon_F 
          \ltk -\frac{1}{2} \left[ 1 -\log{\left( \frac{\tilde{\Delta}_B^2+\tdf^2}{4} \right)}
                            \right] 
          \lk \tilde{\Delta}_B^2 
              +\tdf^2 \rk  
                               +\frac{2}{2\zeta+|\xi|} \tilde{\Delta}_B^2
                               +\frac{2}{|\xi|} \tdf^2 \rtk.  
\label{eED2}
\end{eqnarray} 
Fig.~\ref{Figc} shows a contour plot of the energy density 
in $\tilde{\Delta}_B$-$\tdf$ plane 
when $\zeta=0.5$ and $\xi=-0.6$. 

As is clear from eq.~(\ref{eED2}), 
the coefficient of $\tilde{\Delta}_B^2$ is larger than that of $\tdf^2$, 
$2/(2\zeta+|\xi|) > 2/|\xi|$, 
so that the minimum point of the energy density is always 
on the $\tilde{\Delta}_B$ axis. 

\begin{figure}
\begin{center}
\includegraphics[height=6cm]{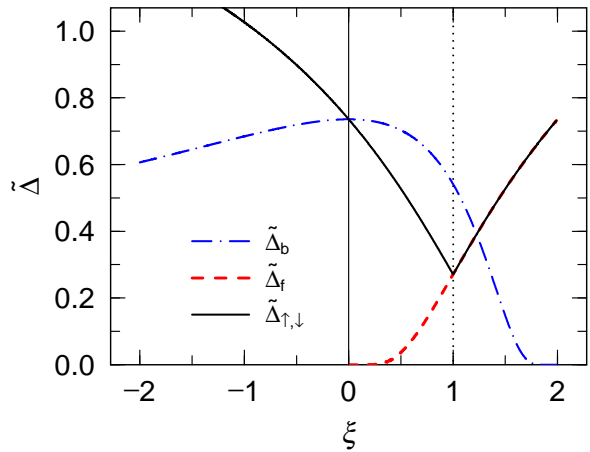}
\end{center}
\caption{The gap functions (extreme values) at $\zeta=1$ as a function of $\xi$. }
\label{op1}
\end{figure}

Thus, for $g_{ff} <0$,
the order parameters take values
$\lk \tdb,\tdf \rk 
     = \lk 2\zeta\frac{2}{2\zeta+|\xi|}e^{\frac{-2}{2\zeta+|\xi|}},0 \rk $,  
and fermion densities become in-phase,  
$\rho_\uparrow(z) = \rho_\downarrow(z)
                  =n_f\ldk 1- \frac{\Delta_b}{2 \epsilon_F \zeta} 
                              \cos(2k_F z) \rdk$, 
which have the same form with those in eq.~(\ref{fden1}) 
in the repulsive fermion-fermion case $g_{ff} >0$.  
The boson condensate has also the same form with (\ref{bden1}).  
However, the $|\xi|$-dependence of the order parameter $\Delta_b$ 
is different in both cases as shown in Fig.~\ref{op1}
where the gap functions are plotted for the effective coupling constant $\xi$ 
when $\zeta=1$. 

\subsection{Phase diagram}

\begin{figure}
\begin{center}
\includegraphics[height=7cm]{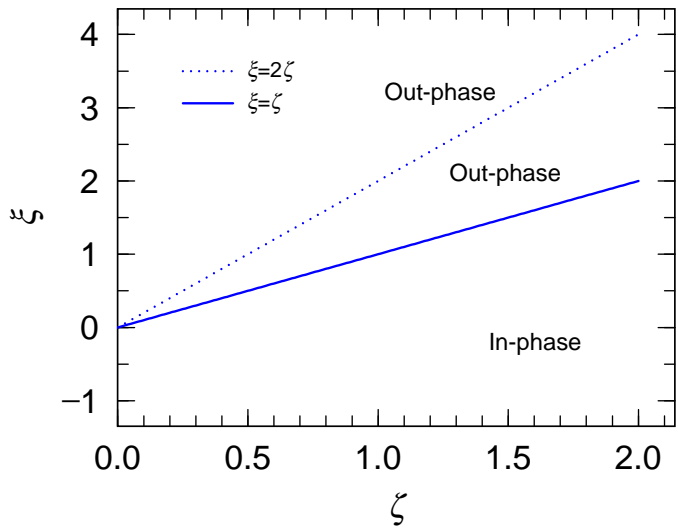}
\end{center}
\caption{The phase diagram of the density-wave structure 
of the ground state in $\zeta$-$\xi$ plane.}
\label{Figb}
\end{figure}

Finally, in Fig.~\ref{Figb}, 
the phase diagram of the density-wave structure in the ground state 
is shown in the plane of the effective coupling constants $\zeta$ and $\xi$ 
in eqs.~(\ref{vzeta},\ref{vxi}), 
which correspond to the $g_{bf}^2/g_{bb}$ and $g_{ff}$. 

In Fig.~\ref{Figb}, 
there exist two phases denoted by in-phase and out-phase, 
which correspond to the fermionic in-phase density wave ($\theta=0$) 
and out-phase density wave ($\theta=\pi$) 
with/without the periodic BEC in the ground state.

In the case of the repulsive fermion-fermion interaction ($\xi>0$), 
there exists a competition between in- and out-phase density waves, 
and the border of these two phases is given by $\xi=\zeta$ 
(rigid line in Fig.~\ref{Figb}). 
Another critical border line $\xi=2\zeta$ exists  
(dotted line in Fig.~\ref{Figb}), and,
above this line, the ground state has the out-phase fermion density waves 
but no periodic BEC. 

As for the attractive fermion-fermion interaction ($\xi<0$), 
the ground state has the periodic BEC 
which is coherent with the in-phase fermion density waves.
The fermion-fermion interaction affects the boson magnitude $\Delta_b$ implicitly. 

The behaviors of the order parameters and the effective gap functions (\ref{egf1}-\ref{egf2}) 
in Fig.~\ref{op1} are reflected 
in the phase diagram;  
$\Delta_f$ and $\Delta_b$ are continuous and smooth functions of $\xi$ 
as stationary points of the energy density, 
and disappear at $\xi=0$ and $\xi=2\zeta$ respectively, as mentioned above.  
The line for the effective gap function $\Delta_{\uparrow, \downarrow}$ 
(solid line),  
the net energy gap at the Fermi surfaces, 
bends at $\xi=\zeta$; 
it implies the transition from the in-phase density wave to the out-phase one.       
      
%

\section{Summary and Discussion} 

We studied the ground state of the Q1D boson-fermion mixture 
of the atomic gas with two-component fermion and one-component boson,  
and showed in the consistently formulated mean-field approximation 
that the states with the periodic BEC and 
the fermion density waves with the wave number $2k_F$ 
become stable therein at $T=0$
We also showed that 
there exists competition 
between the in-phase and out-phase fermion density waves 
in the ground state, 
which is originated from the relative strength 
between the boson-fermion and fermion-fermion effective interactions.
We clarified the wave-density structures of the ground state
and gave the phase diagram of it.

Finally, 
we make a rough estimation for the density-wave amplitudes 
of the atomic gas mixtures in the axially deformed trap potential.    
As possible candidates, 
we take rubidium isotope mixtures,   
${}^{84}{\rm Rb}$-${}^{87}{\rm Rb}$ and ${}^{86}{\rm Rb}$-${}^{87}{\rm Rb}$, 
with the parameters $\omega_a=2 \pi \times 10~{\rm Hz}$ and 
$\omega_r=2 \pi \times 15~{\rm kHz}$ for the trap potential, 
and $N_f\simeq 10^3$ and $N_b\simeq 2 \times 10^4$ for the atom numbers, 
with which the Peierls instability is shown to exist in \cite{MYS1}. 
For the scattering lengths in atomic collisions, 
we take the values in \cite{est1} (atomic unit):  
\begin{eqnarray}
     &&\mbox{1)~${}^{84}{\rm Rb}$-${}^{87}{\rm Rb}$:}\quad
       (a_{ff}, a_{bf}, a_{bb})=(142, 117, 90) 
       \Rightarrow (\zeta, \xi)=(0.397, 0.467),  \\
     &&\mbox{2)~${}^{86}{\rm Rb}$-${}^{87}{\rm Rb}$:}\quad 
       (a_{ff}, a_{bf}, a_{bb})=(7, 336, 90) 
       \Rightarrow (\zeta, \xi)=(3.276, 0.023). 
\end{eqnarray}
Using these parameters, 
we obtain the results that, 
in ${}^{84}{\rm Rb}$-${}^{87}{\rm Rb}$ mixture (the case 1)), 
the out-phase density wave should occur with the amplitude 
$\Delta_f/(\epsilon_F \xi) \simeq 0.118$, 
and, in ${}^{86}{\rm Rb}$-${}^{87}{\rm Rb}$ mixture (the case 2)), 
the ground state has the in-phase one with the amplitude 
$\Delta_f/(2\epsilon_F \zeta) \simeq 0.451$
and periodic BEC with amplitudes $\Delta_b/(g_{bf} n_b) \simeq 0.321$. 
The changes of the atom number, the trap potential size 
and also the interaction strengths should make the shift of 
these results, and make the phase transitions shown in Fig.~6 
expected to be observable.  

%

\begin{appendix}

\section{Diagonalization of the mean-field Hamiltonian}
In this appendix, 
we briefly sketch the diagonalization 
of the mean-field Hamiltonian (\ref{MFH1}) 
under the periodic BEC and the fermion density waves 
given in eqs.~(\ref{BD1}) and (\ref{FD2}), 
and show the quasi-particle wave functions.  

The up- and down-fermion contributions in eq.~(\ref{MFH1}) 
have the same form in structure 
and the calculations go completely in parallel, 
so that we omit the subscript of the fermion field operator 
in this appendix.
The first and second terms in eq.~(\ref{MFH1}) should be
diagonalized by a unitary matrix $U$:
\begin{eqnarray}
     && \left(\begin{array}{c}  
              \psi(k) \\
              \psi(k+2k_F)
              \end{array}\right)^\dagger
        U U^{-1}
       \left[ \begin{array}{c c}
              \epsilon(k) & \Delta    \\
              \Delta^*   &  \epsilon(k+2k_F)
              \end{array} \right]
        U U^{-1}
       \left(\begin{array}{c}
             \psi(k) \\
             \psi(k+2k_F)
             \end{array}\right) \nn
    &&\qquad\qquad
      = \left(\begin{array}{c}
              {\alpha^-(k)} \\
              {\alpha^+(k)}
              \end{array}\right)^\dagger 
        \left[\begin{array}{c c}
              E^-(k) & 0   \\
              0  &  E^+(k)
              \end{array}\right] 
        \left(\begin{array}{c}
              \alpha^-(k) \\
              \alpha^+(k)
              \end{array}\right),  
\end{eqnarray}
where $E^\pm(k)$ are the quasi-particle spectrum 
given in eqs.~(\ref{qpe1}) and (\ref{qpe2}), 
and $\alpha^\pm(k)$ is the field operator 
corresponding to them. 

The explicit form of the unitary matrix is given by 
\begin{equation}
     U=\left[ \begin{array}{c c}
                 \tilde{v}(k) & \tilde{u}(k)    \\
                 u(k) & v(k)
                 \end{array}\right], \qquad
     U^{-1}=U^\dagger=\left[ \begin{array}{c c}
                               \tilde{v}^*(k) &  u^*(k)    \\
                               \tilde{u}^*(k)  & v^*(k)
                               \end{array}\right],
\end{equation}
where
\begin{eqnarray}
     && \tilde{u}(k) =\frac{\Delta}{|\Delta|} \sqrt{\frac{1}{2}} 
            \ldk 1 +\frac{\epsilon(k)
                  -\epsilon(k+2k_F)}{ \sqrt{\ltk \epsilon(k)
                                                -\epsilon(k+2k_F) \rtk^2
                                     +|\Delta|^2}} \rdk^{1/2} \\
     && \tilde{v}(k) =-\frac{\Delta}{|\Delta|} \sqrt{\frac{1}{2}} 
            \ldk 1 -\frac{\epsilon(k)
                  -\epsilon(k+2k_F)}{\sqrt{\ltk \epsilon(k)
                                               -\epsilon(k+2k_F) \rtk^2
                                     +|\Delta|^2}} \rdk^{1/2} \\
     && u(k) =\sqrt{\frac{1}{2}} 
              \ldk 1 +\frac{\epsilon(k)
                   -\epsilon(k+2k_F)}{\sqrt{\ltk \epsilon(k)
                                                -\epsilon(k+2k_F) \rtk^2
                                      +|\Delta|^2}} \rdk^{1/2} \\
     && v(k) =\sqrt{\frac{1}{2}} 
              \ldk 1-\frac{\epsilon(k) 
                   -\epsilon(k+2k_F)}{\sqrt{\ltk \epsilon(k)
                                                -\epsilon(k+2k_F)\rtk^2
                                      +|\Delta|^2}} \rdk^{1/2}, 
\end{eqnarray}
for $-k_F\le k \le 0$. 
The quasi-particle wave functions, $u(k)$ and $v(k)$, 
have been normalized to satisfy the condition:
$|u(k)|^2+|v(k)|^2=|\tilde{u}(k)|^2+|\tilde{v}(k)|^2=1$. 

\section{Gap equations}

The gap equations of the system are given by
\begin{eqnarray} 
     &&-\frac{4}{g_{ff}} \ltk \Delta_f \cos(\theta)+
                              \Delta_b' \cos\left(\frac{\theta}{2} \right) 
                                        \cos\left(\varphi+\frac{\theta}{2} \right) 
                         \rtk \nn
     &&\qquad= 2 \int_{-k_F}^0 \frac{dk}{2\pi}
           \ltk \frac{(\Delta_b+\Delta_b') \cos(\varphi) 
                      +\Delta_f}{\sqrt{ \ltk \epsilon(k)
                                            -\epsilon(k+2k_F)\rtk^2
                                            +|\Delta_\uparrow|^2 }}
      +\frac{(\Delta_b+\Delta_b') \cos(\varphi+\theta)
            +\Delta_f}{\sqrt{ \ltk \epsilon(k)
                                  -\epsilon(k+2k_F) \rtk^2
                             +|\Delta_\downarrow|^2 }} \rtk  \\
{}\nn
     &&\frac{2\Delta_b}{G_{bf}}  
     = 2 \int_{-k_F}^0 \frac{dk}{2\pi} 
            \ltk \frac{\Delta_f \cos(\varphi) 
                      +\Delta_b
                      +\Delta_b'}{\sqrt{ \ltk \epsilon(k)
                                             -\epsilon(k+2k_F)\rtk^2
                                        +|\Delta_\uparrow|^2 }}
      +\frac{\Delta_f \cos(\varphi+\theta)
            +\Delta_b
            +\Delta_b'}{\sqrt{ \ltk \epsilon(k)
                                   -\epsilon(k+2k_F) \rtk^2
                              +|\Delta_\downarrow|^2 }} \rtk \\ 
{}\nn
     &&-\frac{4}{g_{ff}} \ltk \Delta_b'
                             +\Delta_f \cos\left(\frac{\theta}{2} \right) 
                                       \cos\left(\varphi+\frac{\theta}{2} \right) 
                         \rtk  \nn 
      &&\qquad= 2 \int_{-k_F}^0 \frac{dk}{2\pi} 
            \ltk \frac{\Delta_f \cos(\varphi)
                      +\Delta_b
                      +\Delta_b'}{\sqrt{ \ltk \epsilon(k)
                                             -\epsilon(k+2k_F)\rtk^2
                                             +|\Delta_\uparrow|^2 }}
       +\frac{\Delta_f \cos(\varphi+\theta)
             +\Delta_b
             +\Delta_b'}{\sqrt{ \ltk \epsilon(k)
                                    -\epsilon(k+2k_F) \rtk^2
                               +|\Delta_\downarrow|^2 }}\rtk 
\end{eqnarray}
which can be derived from the energy density eq.~(\ref{ED1}) 
by taking the variation with respect to the gap functions 
$\Delta_f$, $\Delta_b$, and $\Delta_b'$ each other.

\end{appendix}

\end{document}